\documentclass[12pt]{article}
\newcommand{\be}{\begin{eqnarray}}
\newcommand{\ee}{\end{eqnarray}}

\usepackage{amsmath}
\usepackage{amssymb}
\usepackage{epsfig}

\begin{document}

\title{\hfill{\tiny FZJ--IKP(TH)--2004--06, HISKP-TH-04/13} \\[1.8em]  On the determination of the parity of 
the $\Theta^+$}
 
\author{C. Hanhart$^1$, J. Haidenbauer$^1$, K. Nakayama$^{1,2}$, U.-G.
  Mei{\ss}ner$^{1,3}$\\
{\small $^1$Institut f\"{u}r Kernphysik, Forschungszentrum J\"{u}lich GmbH,}\\ 
{\small D--52425 J\"{u}lich, Germany} \\
{\small $^2$ Dept. of Physics and Astronomy, University of Georgia,}\\ 
{\small Athens, Georgia 30602, USA}\\ 
{\small $^3$Helmholtz-Institut f\"{u}r Strahlen- und Kernphysik (Theorie), 
Universit\"at Bonn}\\ 
{\small Nu{\ss}allee 14-16, D--53115 Bonn, Germany}}

\maketitle

\begin{abstract}
\noindent We critically examine
the possibility of determining the parity of the $\Theta^+(1540)$ from the
reactions $NN\to \Theta^+Y$ ($Y$ = $\Lambda$, $\Sigma$) 
recently discussed in the literature.  Specifically, we study the
energy dependence of those observables that were suggested to be
the most promising ones to unravel the parity of the $\Theta^+$, namely
the spin correlation coefficient $A_{xx}$, and the spin transfer coefficient 
$D_{xx}$. 
We show that the energy dependence of $\sigma_0(1+A_{xx})$, 
corresponding to the spin-triplet production cross section, guarantees 
unambiguous information on the parity of the $\Theta^+$. Here, $\sigma_0$ 
denotes the unpolarized cross section. Also, the possibility to determine the
parity of the $\Theta^+$ through the energy dependence of $\sigma_0D_{xx}$
is discussed.
\end{abstract}

\newpage

\pagestyle{plain}


\section{Introduction}
By now there is increasing an experimental evidence for the existence of an
exotic baryon state, the $\Theta^+(1540)$ pentaquark \cite{review}. 
Its parity, however, is not yet identified. This quantum number is a decisive 
quantity regarding the substructure of the observed strangeness $S=+1$ 
resonance \cite{Jennings}. In recent months
several proposals were put forward to determine its parity, in both hadronically 
\cite{MODELS1,thomas,unsers,nam,nam2,uzikov1,uzikov2} and electromagnetically 
\cite{MODELS2,NL} induced reactions. In these 
reactions, both the model-dependent and model-independent analyses reveal that 
certain spin observables are directly related and/or very sensitive to the parity 
of the $\Theta^+$. 
Indeed, based on reflection symmetry in the scattering plane, it has been 
shown in Ref. \cite{NL} that it is straightforward to identify the spin observables 
which are directly related to the parity of the $\Theta^+$. Although the considerations
in that work have been confined to the photoproduction reaction, the method
discussed there for the parity determination is quite general and can be easily applied 
to other reactions induced either by hadronic or electromagnetic probes. However, the 
method requires the polarization of the $\Theta^+$ to be measured, a requirement 
that poses an enormous experimental challenge.    
  
Amongst the existing proposals, the one by A. Thomas et al. \cite{thomas} to 
measure $\vec p\vec p\to
\Theta^+\Sigma^+$ seems to be most appealing, for it points to a
model-independent determination of the parity without the need to determine
the $\Theta^+$ polarization. 
Moreover, the first observation of the  $\Theta^+$ in an nucleon--nucleon ($NN$) collision \cite{Eyrich}
suggests that the production cross section is in the order
of 0.4~$\mu b$ so that concrete experimental investigations on the $\Theta^+$
parity in this reaction seem to be indeed feasible. 
The idea by A. Thomas et al. exploits the fact that in the $NN$ system the
Pauli principle links spin and parity: In order to obey Fermi statistics the
quantum numbers characterizing the $NN$ system have to fulfill
\begin{equation}(-)^{L+S+T}=-1 \ , \label{pauli}\end{equation}
where $L$, $S$ and $T$ are the orbital angular momentum and the total 
spin and isospin, respectively. 
Thus---for a given isospin---the total parity $\pi$ of the system is closely linked
to its spin, since $\pi = (-)^L$. If the final state is in an $s$--wave we
have, in addition, $\pi = \pi{(\Theta^+)}$, which allows to determine the parity of the
exotic baryon through a manipulation of the initial spin state.    
In Ref. \cite{unsers} this proposal was worked out in detail and, in particular, 
the spin correlation parameter $A_{xx}$, was identified as {\it the}
crucial observable and its energy dependence was discussed based on general 
arguments. The findings of Ref. \cite{unsers} were shortly afterwards supported
by a particular model calculation \cite{nam}. 

A very important observation was made in Refs. \cite{uzikov1}, namely,
that if the $\Theta^+$ is an isoscalar then the channel $pn\to \Theta^+\Lambda$
offers independent additional information. As  should be clear from
Eq. (\ref{pauli}), the role of the positive and negative parities is
interchanged when switching from isospin $T=1$ to $T=0$.
In addition, other observables and the case of arbitrary spin of 
the $\Theta^+$ were discussed in Refs. \cite{uzikov2}. 
In particular, it was argued that the spin transfer coefficient $D_{xx}$ 
plays a special role amongst all possible polarization observables that can be measured in
$NN$ induced reactions, for it might allow a determination of the parity of the $\Theta^+$
with currently available experimental facilities. This is because the hyperons 
have a self-analyzing decay that allows to determine the hyperon polarization 
solely from the angular pattern of the decay particles.
So far these considerations have been 
restricted to threshold kinematics only. 

In this paper we re-examine the suggested methods
for a parity determination in $NN$ collisions. Furthermore, 
we provide detailed information for an explicit experimental determination of the parity of the $\Theta^+$. 
Specifically, we
shall 
\begin{itemize}
\item show that the energy dependence of $\sigma_0(1+A_{xx})$, where $\sigma_0$ denotes the
  unpolarized cross section, leads to an unambiguous determination of the parity of the
  $\Theta^+$\, ; 
\item discuss the possibilities to determine the parity from a measurement of
  the spin transfer coefficient $D_{xx}$.
\end{itemize}

The above two points will be examined on general grounds. Furthermore, in order to
corroborate our findings, we complement them 
with a concrete calculation within a meson--exchange model.
To be concrete we consider the case of a spin $1/2$--$\Theta^+$ only.

\section{The ideal observable}
\begin{figure}[t]
\begin{center}
\epsfig{file=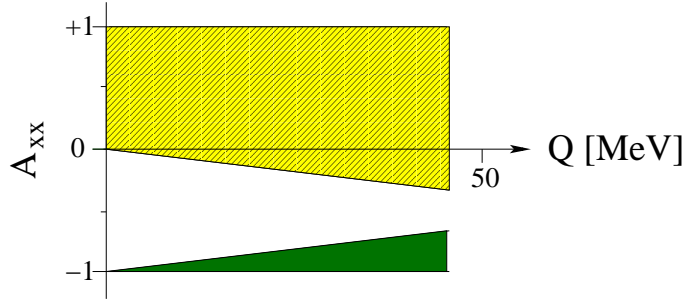, height=4cm}
\caption{\it Sketch of the expected energy dependence of $A_{xx}$ for the two
  different parities of the $\Theta^+$~\protect\cite{unsers}. In case of the 
  $pp \to \Theta^+\Sigma^+$ ($pn \to \Theta^+\Lambda$) reaction
  the hatched area corresponds to negative (positive) parity, whereas the
  filled area corresponds to positive (negative) parity. }
\label{axx_mi}
\end{center}
\end{figure}

In
Ref. \cite{unsers} it was shown that
\begin{eqnarray}
^3\sigma_\Sigma=\frac12\sigma_0(2+A_{xx}+A_{yy}) 
\label{asigdef}
\end{eqnarray}
projects on spin triplet initial states in the $NN\to\Theta^+Y$ reaction
($Y=\Sigma,\Lambda$). From this, it was concluded \cite{unsers} that a 
measurement of $A_{xx}$ in these reactions would be the
ideal quantity for the parity determination of the $\Theta^+$. The reason is that,
when approaching the production threshold energy, this quantity has to approach the 
value of -1 in the case of a positive parity pentaquark but necessarily a 
positive value in the case of a negative parity pentaquark in the reaction 
channel $pp\to\Theta^+\Sigma^+$. In the $pn\to\Theta^+\Lambda$ channel, this 
role is interchanged, i.e., $A_{xx}$ approaches -1 in the case of a negative 
parity but a positive value in the case of a positive parity pentaquark.
However, to make these findings of any practical use, it was necessary to 
provide an educated guess for the behavior of $A_{xx}$ away from the physical 
threshold, i.e. for excess energies up to 50 MeV, say, for only there an 
experiment can be performed with
reasonable count rates. In Ref. \cite{unsers} this was done using the so--called
'naturalness assumption': it was assumed that an $l-$th partial wave 
introduces a factor of $(p'/\Lambda)^l$ into the corresponding amplitude, 
where $p'$ is the relative momentum of the 
outgoing particles and $\Lambda$ is a scale typical for the production
itself. Apart from this factor, the (reduced) partial wave amplitudes were assumed 
to be of the same magnitude for all partial waves.
For later reference the corresponding result for the energy dependence of
$A_{xx}$ is shown in Fig. \ref{axx_mi} as a function of the excess energy $Q$. 

\begin{figure}[t!]
\begin{center}
\epsfig{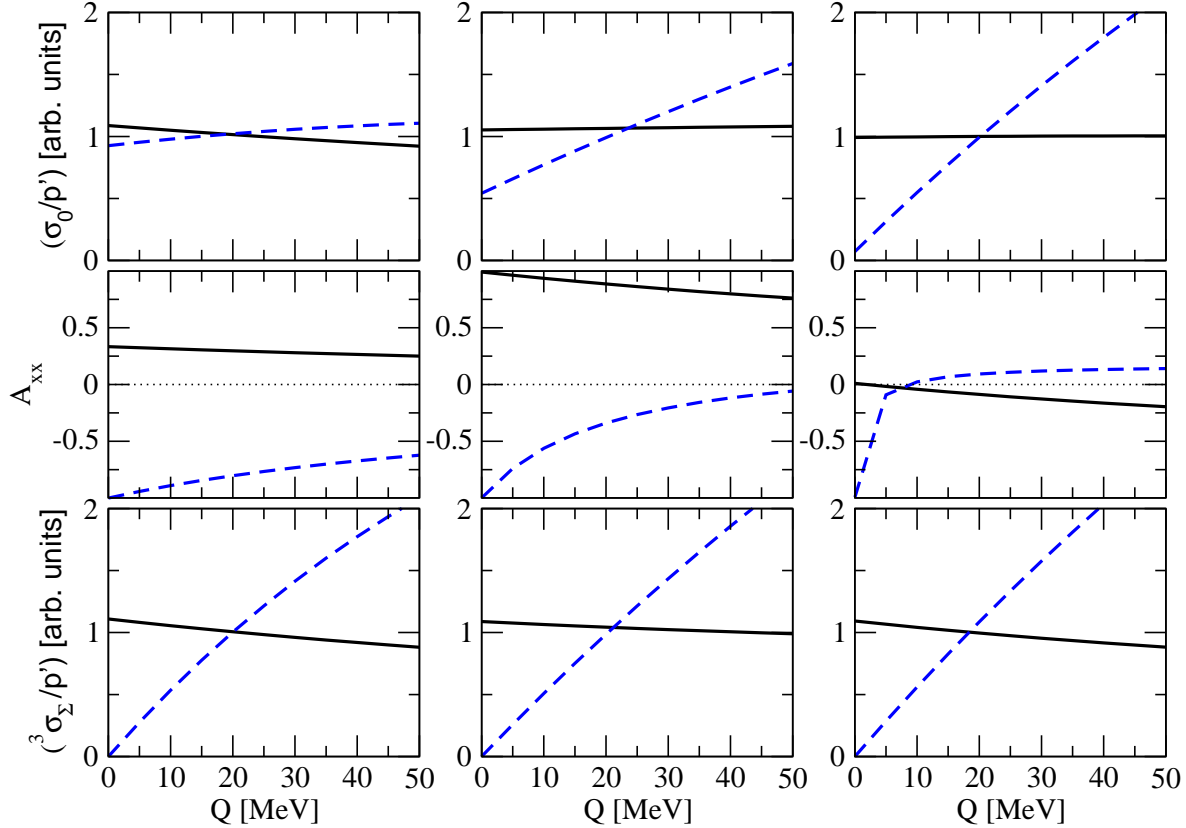}
\caption{\it Energy dependence of the total cross section and the angular
integrated polarization observable $A_{xx}$ and $^3\sigma_\Sigma$
for the reaction $pp\to \Sigma^+\Theta^+$.
Solid (dashed)
lines correspond to a negative (positive) parity $\Theta^+$. Shown are
results for 
three different models for the production operator: the left column shows the
results for the model with only kaon exchange, the middle one those for the
one with $K^*$ exchange and the right one those for the model including $K^*$ 
and $K$ exchange. All results for $\sigma_0$ and $^3\sigma_\Sigma$ are 
normalized to 1 at an excess energy of 20 MeV and are
divided by the phase-space volume. 
}
\label{pp}
\end{center}
\end{figure}

Naturally the question arises how reliable are the scale arguments mentioned above.
Final state interactions can introduce additional large scales into the problem. 
However, as was argued in Ref. \cite{unsers}, there are good reasons
to believe that the $\Theta^+$--hyperon interaction is weak. It is also
straightforward to show that the Coulomb interaction has no influence anymore
for excess energies of 10 MeV or higher. However, scale arguments typically
indicate only the order of magnitude of particular contributions and thus can
well be off by factors of 2--3. In addition, they can not account for possible 
cancellations amongst different production mechanisms. Unfortunately, these
uncertainties can lead to wrong conclusions on the parity of the $\Theta^+$,
because they could change significantly the energy dependence of $A_{xx}$
compared to the one shown in Fig. \ref{axx_mi}. For illustration of this point
in the middle row of Figs. \ref{pp} and \ref{pn} we show the results for $A_{xx}$
calculated within various models described in detail in the appendix. A comparison
with the prediction in Fig.~\ref{axx_mi} shows that $A_{xx}$ in case of a
positive parity pentaquark (dashed curves) in 
the second and third columns in Fig.~\ref{pp} would lead to an erroneous assignment 
of the parity of the pentaquark. The same can be said about the negative
parity results for the $pn$ reaction, shown as the solid curve
in the third column of Fig.~\ref{pn}.   
It is thus necessary
to look for observables that allow solid conclusions about the parity of the
resonance produced irrespective of the relative importance of various
amplitudes.

\begin{figure}[t!]
\begin{center}
\epsfig{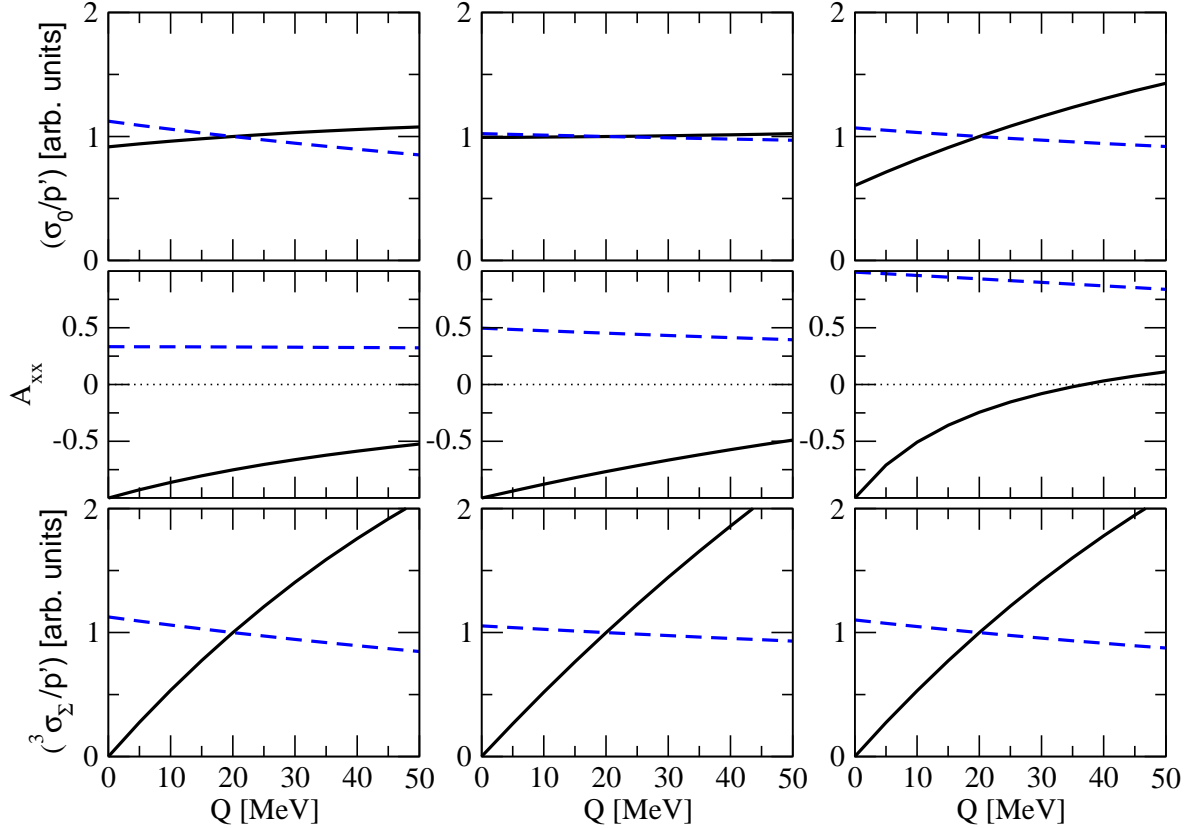}
\caption{\it Energy dependence of the total cross section and the angular
integrated polarization observable $A_{xx}$ and $^1\sigma_\Sigma$ 
for the reaction $pn\to \Lambda\Theta^+$. 
Same description of curves and panels 
as in Fig. \protect\ref{pp}. All results for $\sigma_0$ and
$^3\sigma_\Sigma$ are normalized to 1 at an excess energy of 20 MeV and are
divided by the phase-space volume.
}
\label{pn}
\end{center}
\end{figure}

Such observables indeed exist! The key observation is that
$^3\sigma_\Sigma$ projects onto the spin triplet initial states and, therefore, its
energy dependence will unambiguously point to the parity of the pentaquark.
Let us discuss this argument in more detail: we consider first the 
$pp\to \Sigma^+\Theta^+$ reaction. 
Odd (even) partial waves in the final state correspond to odd (even) partial waves in the 
initial state in case of a positive parity pentaquark but to 
even (odd) partial waves in the initial state for a negative parity pentaquark. 
From Eq. (\ref{pauli}) we find that for this reaction the initial state
is in a spin triplet (singlet) state for positive parity but in a spin singlet (triplet) 
state for negative parity. Thus, the lowest (even) partial wave in the final
state corresponds to a spin singlet or triplet initial state for positive or
negative parity, respectively. Since the centrifugal barrier leads to a momentum
dependence of the $l-$th partial wave amplitude of $p'\, ^{(l+1)}$  
we find that the spin triplet cross section in the $pp$ induced reaction 
should scale as 
 $\sqrt{Q}$ for a negative parity pentaquark and as
$Q^{3/2}$ for a positiv parity pentaquark\footnote{
The individual production mechanisms will induce an additional weak energy
dependence on top of this. We do not expect the production operator to introduce 
any stronger energy dependence since it is of rather short-ranged nature. 
In fact, the momentum transfer at threshold is given by 
$q = \sqrt{(m_\Theta + m_\Sigma)^2/4 - m_N^2} \simeq 5 \, \rm{fm}^{-1}$ which 
corresponds to a range of $1/q \simeq 0.2 \rm{fm}$.}. 
This observation on the energy dependence of the spin triplet cross section holds
irrespective of the relative importance of different partial waves.
For illustration, in Fig. \ref{pp} we show the results of various model
calculations (see Appendix for details) for the energy dependence of the
total cross section $\sigma_0$, $A_{xx}$ and the spin triplet cross
section $^3\sigma_\Sigma$. 
(Note that the results shown in Figs. \ref{pp} and \ref{pn} are 
divided by the factor $\sqrt{Q}\propto p'$, i.e. by the two-body phase-space
volume.)
The results for the positive parity pentaquark are shown as
dashed lines whereas those for the negative parity pentaquark are shown as
solid lines. In order to exhibit better the difference in the energy dependence of the
observables corresponding to different parities of the pentaquark, the results
for $\sigma_0$ and $^3\sigma_\Sigma$ are scaled to 1 at $Q=20$ MeV for all models.
Although the energy dependence of $A_{xx}$ exhibited by two of the models 
considered in Fig. \ref{pp} is significantly stronger than what is expected for natural-size
amplitudes (c.f. Fig. \ref{axx_mi}), the energy dependence of
$^3\sigma_\Sigma$ for the two parities always follows the features discussed above 
and, consequently, allows definite conclusions on the parity of the pentaquark.

Analogously, for the $pn\to \Lambda\Theta^+$ reaction the spin triplet
cross section should scale as $Q^{3/2}$ for odd parity pentaquarks and as
$\sqrt{Q}$ for even parity pentaquarks. In Fig. \ref{pn} we show the results
of the same models used before for the $pn$ channel. Again, we
observe that the scaling argument leading to the results of $A_{xx}$ presented 
in Fig. \ref{axx_mi} is violated here for one of the models 
considered, but the energy dependence of $^3\sigma_\Sigma$
allows definite conclusions on the parity of the pentaquark.

One might still ask what would happen if for some reason the $s$--wave 
amplitude is strongly suppressed or even absent. Even in this case the
method of parity determination proposed will work. To be concrete let us look
at the channel $pp\to \Sigma^+\Theta^+$ only. If the pentaquark has positive
parity a spin triplet initial state is to be at least in a $p$--wave.  Thus,
$^3\sigma_\Sigma$ should follow a $Q^{3/2}$ behavior independent of the
strength of the $s$--wave \footnote{We tuned the parameters of the $K-K^*$ model such
that this case is realized in the right column of Fig. \ref{pn}.}. If the
pentaquark has negative parity, spin triplet initial states lead to even partial wave
final states. In the absence of an $s$--wave this implies a behavior at least
as $Q^{5/2}$ for $^3\sigma_\Sigma$. Consequently we should expect the spin
triplet cross section to be strongly suppressed in comparsion to the total
cross section which should show a $Q^{3/2}$ behavior\footnote{We were not able to construct a
model, where this scenario is realized.}! Again this difference is decisive.

It should also be stressed that all arguments given above hold even if the
$\Theta^+$ signal seen in the experiment would be due to an interference of the
resonance signal with the background.

Therefore, we propose to measure the total cross section as well as the 
corresponding (angle integrated) spin-correlation parameter $A_{xx}$ for $NN$ 
induced pentaquark production at two well separated energies,
e.g. at excess energies of $Q=20$ and 40 MeV. 

\section{The spin transfer coefficient $D_{xx}$}


In addition to the spin correlation coefficient $A_{xx}$ also the spin
transfer coefficient $D_{xx}$ is discussed in the literature as a useful
observable to determine the parity of the pentaquark in $NN$ collisions. 

\begin{figure}[t]
\begin{center}
\epsfig{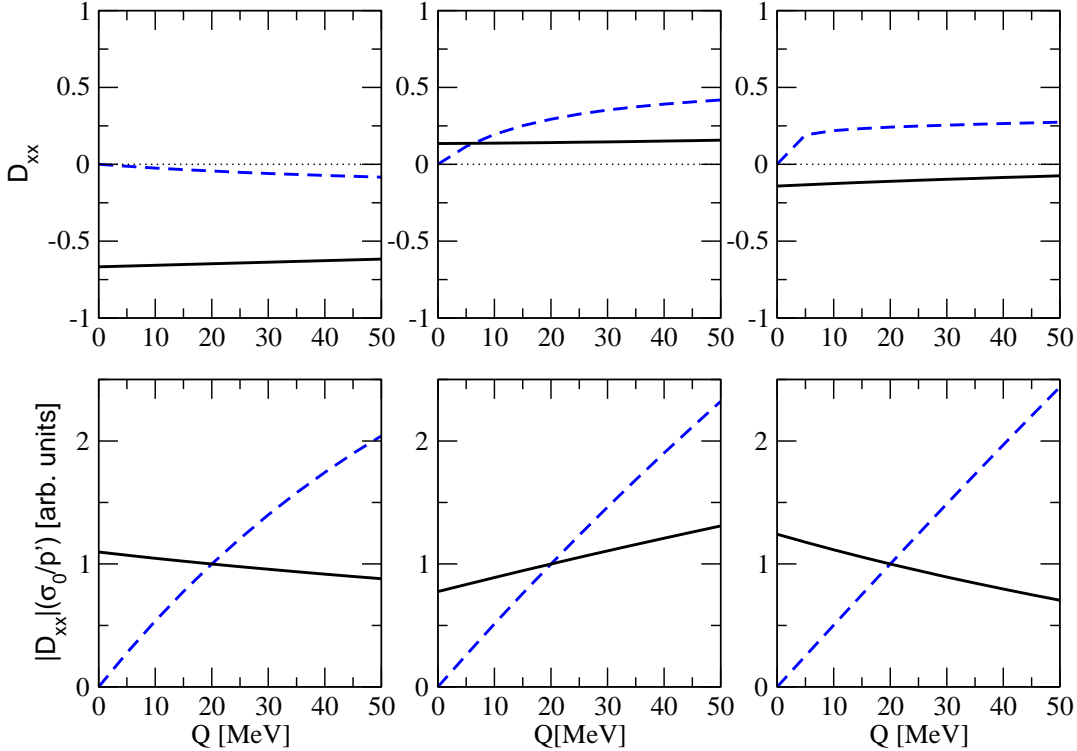}
\caption{\it Results for the energy dependence of  $D_{xx}$ and
  $|D_{xx}|\sigma_0$ in the channel
  $pp\to \Sigma^+\Theta^+$ for the various
  models.  Same description of curves and panels as in Fig. \protect\ref{pp}.}
\label{dxx_pp}
\end{center}
\end{figure}

\begin{figure}[t]
\begin{center}
\epsfig{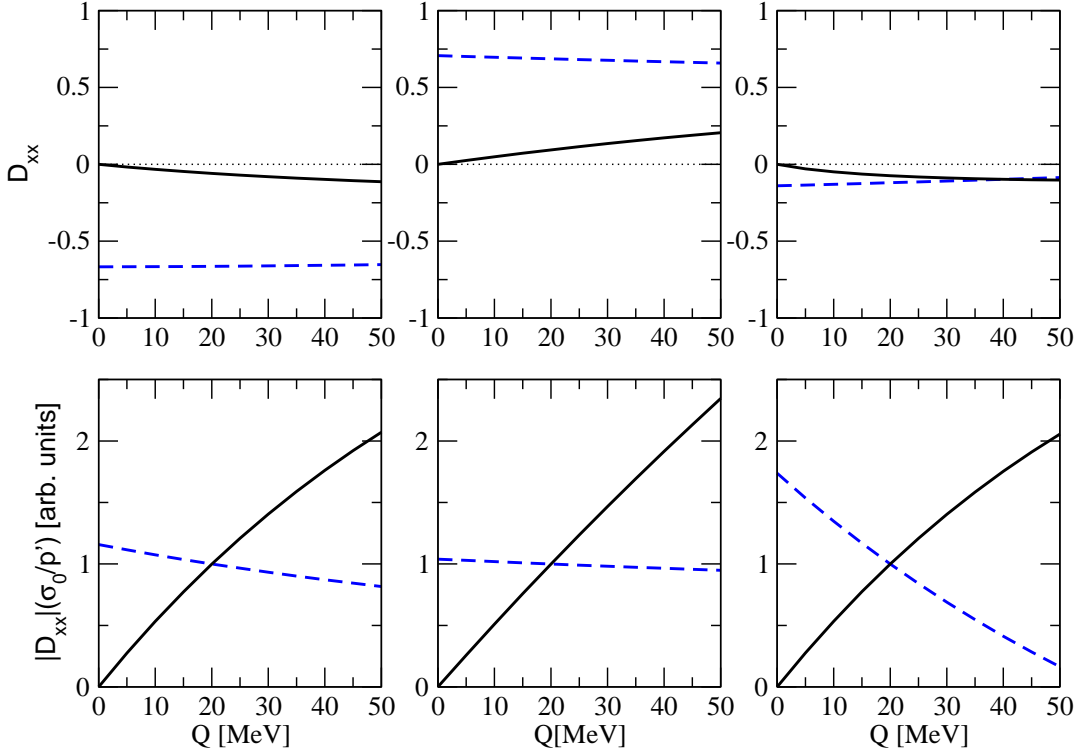}
\caption{\it  Results for the energy dependence of $D_{xx}$ and
  $|D_{xx}|\sigma_0$ in the channel
  $pn\to \Lambda\Theta^+$ for the various models. The calculations for the
  angular distributions were performed for $Q=40$ MeV. The calculations for
  the angular distributions were performed for $Q=40$ MeV.  
  Same description of curves and panels as in Fig. \protect\ref{pp}.}
\label{dxx_pn}
\end{center}
\end{figure}

The energy dependence of $D_{xx}$ cannot be
studied on as general grounds as that of the spin triplet cross section
$^3\sigma_\Sigma$ discussed in the previous section. However, using the
methods outlined in, e.g., Refs. \cite{unsers,report} we may give the general
angular and energy dependence of $D_{xx}$ up to terms of order $(p')^2$. This
analysis leads to the following structure for negative parity pentaquarks produced
in the $pp \to \Theta^+\Sigma^+$ reaction,
\begin{equation}
4\sigma_0D_{xx}=\alpha+\left(\frac{p'}{\Lambda}\right)\beta\cos(\theta)+\left(\frac{p'}{\Lambda}\right)^2
\left[\gamma\cos^2(\theta)+\delta\sin^2(\theta)\sin^2(\phi)\right] + 
{\cal O}\left(\frac{p'^3}{\Lambda^3}\right) \ ,
\label{oddpp}
\end{equation}
where all the coefficients ($\alpha,\beta,...$) turn out to be independent 
from each other. On the other hand, for positive parity pentaquarks produced in
the $pp$ induced reaction we find
\begin{equation}
4\sigma_0D_{xx}=\left(\frac{p'}{\Lambda}\right)\beta'\cos(\theta)+\left(\frac{p'}{\Lambda}\right)^2
\left[\epsilon'+\gamma'\cos^2(\theta)+\delta'\sin^2(\theta)\sin^2(\phi)\right] + 
{\cal O}\left(\frac{p'^3}{\Lambda^3}\right) \ ,
\label{evenpp}
\end{equation}
where again all the coefficients ($\beta',\gamma' ...$) turn out to be
independent from each other. Eqs. (\ref{oddpp}) and (\ref{evenpp}) reflect the
different threshold behavior for the two different parities first pointed out in
Refs. \cite{uzikov1}: only for negative parity pentaquarks produced in the
$pp$ reaction a non--vanishing threshold value is allowed for
$D_{xx}$. However, nothing prevents $\alpha$ from being small and thus it can
not be said a priori, whether a conclusion on the parity of the pentaquark can be
drawn from a measurement of $D_{xx}$---we come back to this issue below. The
relevant expressions for the $pn \to \Theta^+\Lambda$ channel are identical to Eqs. (\ref{oddpp})
and (\ref{evenpp}), except that the roles of the different parities are interchanged.

To simplify the discussion we will now focus on the $pp$ system only.
The terms higher order in $p'/\Lambda$ 
are given here for the
first time. For both parities the term linear in $p'$ stems
from an interference of $s$-- and $p$--waves in the final state. The terms
quadratic in $p'$ originate from the interference of various spin triplet initial
states and thus stem from $p$--waves interfering with each other in case of an
even parity pentaquark and from
$s$--waves interfering with $d$--waves  in case of an
odd parity pentaquark.

As was stressed in the previous paragraph, we should identify observables that
allow conclusions on the parity without the need to employ scale arguments
for the relative importance of different partial waves. 
It turns out that $D_{xx}\sigma_0$ is such a quantity. 
A close look at Eqs. (\ref{oddpp}) and  (\ref{evenpp}) reveals that ideally
one may extract from the experiment $D_{xx}\sigma_0$ at $\phi=0$ and $\cos
(\theta)=0$ at two different energies, for in this case the production of a negative parity pentaquark
can lead to an energy independent non--zero result (up to the energy dependence from
phase space and small corrections due to the energy dependence of the
production operator), whereas a positive parity
pentaquark can lead to a linear energy dependence (in this case an energy
independent result consistent with zero would not be conclusive, for the parameter $\epsilon'$ can
also be 0). However, this would require a highly differential measurement. 
Fortunately, already the angle integrated result for
$D_{xx}\sigma_0$ can provide valuable information when one extrapolates linearly
from the data to the threshold after dividing by the phase space 
factor $p'$. Here 
a finite value is possible only for a positive parity pentaquark
whereas a negative parity pentaquark has to lead to a vanishing
$D_{xx}\sigma_0$ at threshold. Note, however, that no conclusion on the 
parity can be drawn if this extrapolation leads to a value consistent
with zero (unfortunately we do not
have a model calculation to illustrate this point---see also footnote 3).

In Figs. \ref{dxx_pp} and \ref{dxx_pn} we show results for $D_{xx}\sigma_0$, 
corrected for the phase space behavior, based on the same models used in 
the previous paragraph. As can be seen in
all cases the extrapolation to threshold allows for a discrimination of the
different parities. 

It should be stressed, however, that the experimental uncertainty on 
$D_{xx}\sigma_0$ will be large if the value of $D_{xx}$ is small. 
For example, a small $D_{xx}$ for the $pn\to\Theta^+\Lambda$ channel 
resulted from the considered $K-K^*$ model (right column of Fig. \ref{dxx_pn}).
We can make this statement somewhat more quantitative.
It is well known that the statistical fluctuation in an observable, in this
case $D_{xx}\sigma_0 = [\sigma(++)+\sigma(--)] - [\sigma(+-)+\sigma(-+)]$, 
scales as $\sqrt{N}$, where $N$ denotes the count rate
for the individual spin cross sections. On the other hand, the signal itself
is given by $D_{xx}N$ counts. Thus we find for the relative uncertainty $\delta$ of
$D_{xx}\sigma_0$ 
$$
\delta \simeq \frac{1}{D_{xx}\sqrt{N}} \ .
$$
To reduce the uncertainty in $D_{xx}\sigma_0$ to 10 \%
($\delta = 0.1$) we find a required count rate for the polarized cross section
of $N\simeq 100/D_{xx}^2$, where we assumed 100 \% polarization in the beam
as well as 100 \% acceptance in the final state. More realistic numbers will
increase our count rate estimate. This formula shows that if $D_{xx}$ is
0.5 or larger, one needs at least 400 events in the polarized cross
section. However, if $D_{xx}$ turns out to be only of the order of 0.1 in the energy 
range accessible to the experiment (as it is predicted by
some of the considered models) then one needs $N \ge 10000$. 

To conclude, if it is possible to measure $D_{xx}$ with reasonable accuracy at two
different and sufficiently separated energies, it could be possible to extract the
parity of the pentaquark. If an extrapolation from the measured values
of $D_{xx}\sigma_0$ to the threshold leads to a non--vanishing value in the
$pp \to \Theta^+\Sigma^+$ channel, the parity
of the pentaquark must be negative, whereas if it leads to a non--vanishing
value in the $pn$ induced reaction,
the parity must be positive. It should be stressed, however, that here
significantly higher statistics is necessary compared to the measurement of
$A_{xx}$ for there the energy dependence of $\sigma_0(1+A_{xx})$ was the
quantity of interest whereas here an extrapolation to the threshold is
necessary and for both parities the term linear in energy might be sizable.

\section{Summary}

We examined critically
the possibility to determine the parity of the $\Theta^+(1540)$ from the 
reactions $NN\to \Theta^+Y$, where $Y$ denotes the $\Lambda$ or $\Sigma$ hyperon, 
recently discussed in the literature. Specifically, we studied the energy 
 dependences of the observables that have been suggested in the literature
to be the most promising ones to unravel the parity
of the $\Theta^+$, namely $A_{xx}$ and $D_{xx}$. 
The validity and/or breakdown of general scale arguments were critically examined,
showing that peculiarities of the so far unknown production mechanism of the
$\Theta^+$ could make conclusions based on such scale arguments rather unreliable. 

On the other hand, we showed that 
the energy dependence of $\sigma_0(1+A_{xx})$, corresponding to 
the spin-triplet production cross section, guarantees 
unambiguous information on the parity of
the $\Theta^+$, since it does not rely on any assumptions such as scale arguments.

In addition,
we have demonstrated that the spin transfer coefficient $D_{xx}$ could,
under certain conditions, also be used to determine the parity of the pentaquark.
A non--vanishing value of $\sigma_0D_{xx}$, when extrapolated linearly from the
measured values to the threshold, will tell the parity. However, this 
measurement requires in any case very high statistics. In
addition, only a non--vanishing value at the threshold would be 
conclusive. A threshold value consistent with zero is
in accordance with both parities.

Finally, it should be clear that the presented results are applicable for the 
parity determination of any narrow spin--1/2 baryon resonance in $NN$ collisions, for 
they are based solely on general considerations. 

\section*{Acknowledgment}

We thank F. Rathman and A. Sibirtsev for very useful discussions. The work of
KN is partly supported by COSY grant No. 41445282.

\appendix

\section{Appendix}
\label{models}

\begin{figure}[t]
\begin{center}
\epsfig{file=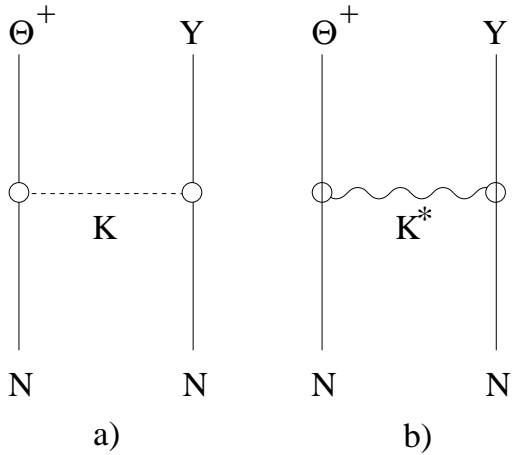, height=6cm}
\caption{\it Diagrams considered in the model calculation.}
\label{diags}
\end{center}
\end{figure}

To illustrate the points made in the main section, we perform a concrete calculation 
within the meson--exchange framework. It should be stressed that this model calculation
is intended to provide only a qualitative picture of the features discussed in work.
For quantitative predictions further and important ingredients need to be incorporated 
into the model. 
For example, inclusion of the
initial state interaction (ISI) would not only change the magnitude of the
total cross sections, but it can also change the relative phase of the amplitudes and
thus change the results of the polarization observables (for a recent
discussion of the effect of the ISI on inelastic $NN$ scattering we
refer to Ref.  \cite{withkanzo}). Also there might be significantly more complicated 
reaction mechanisms that contribute to the production.

So far there is practically no information on the preferred production 
mechanism of the $\Theta^+$. In the only model calculation that has presented 
results for polarization observables \cite{nam}, it was assumed that the
production reaction is dominated by kaon exchange.  However, the apparently rather narrow width of the
$\Theta^+$(1540) \cite{width}, which implies a fairly small $KN\Theta^+$
coupling constant, makes it plausible that other production mechanisms 
should be important if not dominant. One of the obvious candidates is the
exchange of the $K^*$(892) vector meson, whose contribution has been already 
considered in several model calculations. Vector-meson exchanges
yield structures that are similar but of opposite character to those
generated by pseudoscalar mesons (kaons) so that cancellations can occur. 
For example, in the $NN$ system the cancellation between the 
$\pi$- and $\rho$-exchange contributions leads to a significant reduction of the 
tensor force; the same feature is also seen for the $K$- and
$K^*$ exchange in the hyperon-nucleon case \cite{MHE}. 
In the reaction 
$p\bar p \to \Lambda \bar\Lambda$, on the other hand, there is a
strong cancellation between $K$- and $K^*$ exchange in the spin-spin
component, resulting in a strong suppression of the spin-singlet
amplitude as observed experimentally \cite{ppbar}. Thus, it is not
unreasonable to assume that cancellations can also occur in the
$\Theta^+$ production reaction.

Although there might well be more complicated production mechanisms for the
$\Theta^+$ of relevance \cite{Karliner},
we believe that the model of a single kaon exchange (Fig.
\ref{diags}a)), proposed in Ref.  \cite{nn_kex}, or of a single $K^*$ exchange (Fig. \ref{diags}b))
or a combination of both, proposed in Ref. \cite{nn_kandkstex},
is well suited to explore the model dependence of those observables relevant for the
parity determination.  We use the following
Lagrangian densities to calculate the diagrams in Fig. \ref{diags}:
\begin{eqnarray}
{\cal L}^\pm_{NKY} &=& -g_{NKY} 
\bar{N} i \Gamma^\pm Y K  + h.c.,
\nonumber
\\
{\cal L}^\pm_{NK^*Y} &=& -  g_{NK^*Y}
\bar{N} \Gamma^\pm \left[ \gamma_\mu Y 
+ \frac{\kappa }{m_Y + m_N}\sigma_{\mu\nu}Y\partial^\nu
\right]K^{*\mu} + h.c., 
\label{NK*Theta}
\end{eqnarray}
where $\Gamma^+ = \gamma_5$, $\Gamma^- = 1$ and, 
$Y=\Theta^+, \Lambda , \vec\tau\cdot\vec\Sigma$. The superscript $\pm$ refers
to the positive $(+)$ or negative $(-)$ parity of the $\Theta^+$. 
We also introduce a monopole form factor, 
$F(q^2) = (\Lambda_M^2 - m_M^2)/(\Lambda^2_M - q^2)$ ($M=K, K^*$), at each 
vertex, where $q^2$ denotes the squared four--momentum of the exchanged meson $M$
and $m_M$ stands for its mass. The parameter values employed in our calculations are summarized in 
Table \ref{tab1}. Most of them are in the same range as the values
employed/extracted in Refs. \cite{nam,nam2,jul,nij}. The only exception is the
cut--off parameter $\Lambda_{K^*}$. We use $\Lambda_{K^*}=1.3$ GeV instead 
of 1 GeV employed  in Refs. \cite{nam,nam2}, because the latter choice
suppresses the $K^*$ exchange contribution almost completely.
In addition, we varied the $K^*$ tensor coupling in order to exhibit
various scenarios.
The three examples shown in the main text are typical representatives of a large
number of models investigated.

\begin{table*}[t]
\caption{Parameter values used in this work.
All parameters are given in units of $g_{NK\Theta^+}$.
 For the $K^*$, the entries are ($g_{NK^*Y},\kappa$).} 
\begin{center}
\begin{tabular}{l@{\qquad}r@{\qquad}r@{\qquad}r@{\qquad}r}
\hline\hline
   M     & $g_{NM\Theta^+}$ &
 $g_{NM\Sigma}$ & $g_{NM\Lambda}$ & $\Lambda_M$~(MeV) \\
\hline\hline
$K$      &  1         &  3.54         &  -13.26        & 1000    \\
$K^*$    & (0.5, 2)    & (-2.46, -0.5) &  (-5.80, -0.5) & 1300    \\
\hline\hline
\hfill 
\end{tabular}
\label{tab1}
\end{center}
\end{table*}

\end{document}